\documentclass[a4paper,reqno]{amsart}
\usepackage{graphicx}
\usepackage{lscape}
\usepackage[margin=2.5cm]{geometry}
\begin{document}

\newcommand{\beqn}{\begin{equation}}
\newcommand{\eeqn}{\end{equation}}
\newcommand{\beqna}{\begin{eqnarray}}
\newcommand{\ee}{\end{eqnarray}}
\newcommand{\best}{\begin{eqnarray*}}
\newcommand{\ees}{\end{eqnarray*}}

\newcommand{\undertilde}[1]{\underset{\widetilde{}}{#1}}
\newcommand{\al}{\alpha}
\newcommand{\de}{\delta}
\newcommand{\De}{\Delta}
\newcommand{\be}{\beta}
\newcommand{\ga}{\gamma}
\newcommand{\Ga}{\Gamma}
\newcommand{\Be}{\Xi}
\newcommand{\Al}{\Lambda}

\newcommand{\ala}{\bar{\alpha}}
\newcommand{\alb}{\bar{\bar{\alpha}}}
\newcommand{\dea}{\bar{\delta}}
\newcommand{\bea}{\bar{\beta}}
\newcommand{\gaa}{\bar{\gamma}}

\newcommand{\als}{\hat{\alpha}}
\newcommand{\des}{\hat{\delta}}
\newcommand{\bes}{\hat{\beta}}
\newcommand{\gas}{\hat{\gamma}}

\newcommand{\alas}{\hat{\bar{\alpha}}}
\newcommand{\deas}{\hat{\bar{\delta}}}
\newcommand{\beas}{\hat{\bar{\beta}}}
\newcommand{\gaas}{\hat{\bar{\gamma}}}

\newcommand{\as}{\hat{a}}
\newcommand{\ds}{\hat{d}}
\newcommand{\bs}{\hat{b}}
\newcommand{\cs}{\hat{c}}

\newcommand{\ass}{\hat{\hat{a}}}
\newcommand{\dss}{\hat{\hat{d}}}
\newcommand{\bss}{\hat{\hat{b}}}
\newcommand{\css}{\hat{\hat{c}}}

\newcommand{\aas}{\hat{\bar{a}}}
\newcommand{\das}{\hat{\bar{d}}}
\newcommand{\bas}{\hat{\bar{b}}}
\newcommand{\cas}{\hat{\bar{c}}}

\newcommand{\la}{\lambda}
\newcommand{\La}{\Lambda}
\newcommand{\laa}{\bar{\lambda}}
\newcommand{\lab}{\bar{\bar{\lambda}}}
\newcommand{\lac}{\bar{\bar{\bar{\lambda}}}}
\newcommand{\lad}{\overset{4}{\lambda}}
\newcommand{\lae}{\overset{5}{\lambda}}
\newcommand{\laf}{\overset{6}{\lambda}}

\newcommand{\va}{\bar{v}}
\newcommand{\vb}{\bar{\bar{v}}}
\newcommand{\vs}{\hat{v}}
\newcommand{\vas}{\hat{\bar{v}}}

\newcommand{\vo}{v_1}
\newcommand{\voa}{\bar{v}_1}
\newcommand{\vos}{\hat{v}_1}
\newcommand{\voas}{\hat{\bar{v}}_1}

\newcommand{\vt}{v_2}
\newcommand{\vta}{\bar{v}_2}
\newcommand{\vts}{\hat{v}_2}
\newcommand{\vtas}{\hat{\bar{v}}_2}

\newcommand{\vth}{v_3}
\newcommand{\vtha}{\bar{v}_3}
\newcommand{\vths}{\hat{v}_3}
\newcommand{\vthas}{\hat{\bar{v}}_3}

\newcommand{\vf}{v_4}
\newcommand{\vfa}{\bar{v}_4}
\newcommand{\vfs}{\hat{v}_4}
\newcommand{\vfas}{\hat{\bar{v}}_4}

\newcommand{\wa}{\bar{w}}
\newcommand{\wb}{\bar{\bar{w}}}
\newcommand{\wc}{\bar{\bar{\bar{w}}}}

\newcommand{\xa}{\bar{x}}
\newcommand{\xb}{\bar{\bar{x}}}
\newcommand{\xc}{\bar{\bar{\bar{x}}}}
\newcommand{\xd}{\overset{4}{x}}
\newcommand{\xe}{\overset{5}{x}}
\newcommand{\xf}{\overset{6}{x}}

\newcommand{\xs}{\hat{x}}
\newcommand{\xss}{\hat{\hat{x}}}
\newcommand{\xas}{\hat{\bar{x}}}
\newcommand{\xbs}{\hat{\bar{\bar{x}}}}

\newcommand{\ya}{\bar{y}}
\newcommand{\yb}{\bar{\bar{y}}}
\newcommand{\yc}{\bar{\bar{\bar{y}}}}
\newcommand{\yd}{\overset{4}{y}}
\newcommand{\ye}{\overset{5}{y}}
\newcommand{\yf}{\overset{6}{y}}

\newcommand{\ys}{\hat{y}}
\newcommand{\yss}{\hat{\hat{y}}}
\newcommand{\yas}{\hat{\bar{y}}}

\newcommand{\ra}{\bar{r}}
\newcommand{\rb}{\bar{\bar{r}}}
\newcommand{\rc}{\bar{\bar{\bar{r}}}}
\newcommand{\rs}{\hat{r}}
\newcommand{\rss}{\hat{\hat{r}}}
\newcommand{\ras}{\hat{\bar{r}}}

\newcommand{\du}[3]{#1_{#2}^{#3}}
\newcommand{\ol}[1]{\overline{#1}}
\newcommand{\ul}[1]{\underline{#1}}
\newcommand{\os}[2]{\overset{#1}{#2}}

\newcommand\qon{{\rm qP}_{\rm{\scriptstyle I}}}
\newcommand\qtw{{\rm qP}_{\rm{\scriptstyle II}}}
\newcommand\qth{{\rm qP}_{\rm{\scriptstyle III}}}
\newcommand\qfi{{\rm qP}_{\rm{\scriptstyle V}}}
\newcommand\qsi{{\rm qP}_{\rm{\scriptstyle VI}}}

\newcommand\don{{\rm dP}_{\rm{\scriptstyle I}}}
\newcommand\dtw{{\rm dP}_{\rm{\scriptstyle II}}}
\newcommand\dth{{\rm dP}_{\rm{\scriptstyle III}}}

\newcommand\con{{\rm cP}_{\rm{\scriptstyle I}}}
\newcommand\ctw{{\rm cP}_{\rm{\scriptstyle II}}}

\newcommand{\wh}{\widehat}

\newcommand{\eqn}[1]{(\ref{#1})}
\newcommand{\ig}[1]{\includegraphics{#1}}

\newtheorem{theorem}{Theorem}[section]
\newtheorem{lemma}[theorem]{Lemma}
\newtheorem{proposition}[theorem]{Proposition}
\newtheorem{corollary}[theorem]{Corollary}
\newtheorem{conjecture}[theorem]{Conjecture}
\newtheorem{definition}{Definition}[section]
\newtheorem{remark}[theorem]{Remark}
\newtheorem{example}[theorem]{Example}

\title[Fake Lax pairs]{Simple identification of fake Lax pairs}
\author{Samuel Butler and Mike Hay}
\address{INFN, Roma Tre University, Department of Mathematics and Physics}
\email{hay.michael.c@gmail.com}

\begin{abstract}
Two simple ways to identify and explain fake Lax pairs are provided. The two methods are complementary, one involves finding a gauge transformation which can be used to remove the associated nonlinear system's dependent variable(s) from a fake Lax pair. The second method shows that excess degrees of freedom exist in fake Lax pairs. We provide several examples to illustrate both tests. The tests proposed here can be applied to all types of Lax pairs, including scalar or matrix linear problems associated with ordinary or partial difference and/or differential systems.
\end{abstract}
\maketitle

\section{Introduction}
The existence of fake Lax pairs is well known but the phenomenon is not widely understood.

The term Lax pair can refer to several different types of systems, the aspect these all have in common is that a pair of linear equations (an overdetermined system) is associated with a nonlinear integrable system through a compatibility condition. The most important property of a real Lax pair is that it proves the integrability of the associated nonlinear system. Indeed, real Lax pairs are very useful for finding information about the solutions to nonlinear integrable systems. 

Fake Lax pairs, on the other hand, say nothing about the integrability of the associated nonlinear system. Fake Lax pairs often appear very similar to their real counterparts and experts in the area of integrable systems continue to inadvertently publish fake Lax pairs that they believe are real (see sections \ref{sec:hone} and \ref{sec:gramani} for examples). As such, there is serious need for a straightforward method to distinguish between real and fake Lax pairs.

In this letter, we provide two very simple methods to identify fake Lax pairs. Lax pairs can be discrete or continuous, matrix or scalar, used for inverse scattering or isomonodromy, and fake Lax pairs reside in all of these categories. The methods to identify fake Lax pairs presented in this article is applicable to any kind of Lax pair that contains a non-removable spectral variable. The question of inserting a non-removable spectral variable into a Lax pair that does not have one is explored in \cite{m10}.

The methods outlined in this article highlight properties that are sufficient for a Lax pair to be fake, but are not necessary. It is possible that a Lax pair could pass these tests and still be fake, however, every fake Lax pair known to the authors fails both tests.

Although there are many references to fake Lax pairs in the literature, the most famous being \cite{cn91}, there are fewer articles that set out to explain what fake Lax pairs are and how to identify them. So far the methods given to identify fake Lax pairs \cite{lsz90, lll10, m04, s01, s02} have been limited in their applicability and often difficult to apply. In contrast, the tests given here are easily comprehensible and widely applicable.

This article is organised as follows. In section \ref{sec:id} we briefly describe two  methods to identify fake Lax pairs. The methods are best understood using examples so in section \ref{sec:examples} we give several of those taken from the literature. The article terminates with a conclusion in section \ref{sec:conclusion}.


\section{Two methods to identify fake Lax pairs}\label{sec:id}

In this section we describe the methods used to identify fake Lax pairs (FLPs). The use of these methods is illustrated by way of examples in section \ref{sec:examples} below. 

\subsection{Method by removal of the dependent variable}\label{sec:gauge}

In all FLPs found we have been able to use gauge transformations to remove all dependent variables in the associated nonlinear system from the Lax equations. 

\begin{definition}\label{defFLP}
A Lax pair is called {\bf g-fake} if, on solutions of the equation appearing in its compatibility condition, one can remove all dependent variables in the associated nonlinear system from the Lax pair by applying gauge transformations.
\end{definition}

Definition \eqref{defFLP} applies to Lax pairs of any form. To illustrate how it works we consider an $N\times N$ discrete Lax pair of the form
\beqn\label{LP1}
\ol{\phi}=L\phi, \qquad \wh{\phi} = M\phi,
\eeqn
where $\ol{\phi}$ and $\wh{\phi}$ denote iterations of $\phi$ in two discrete independent variables. Suppose the entries of $L$ and $M$ depend on the dependent variable $u$, say, but are otherwise autonomous, and the compatibility condition $\wh{L}M=\ol{M}L$ of this Lax pair is the (nonlinear) equation $Q(u,\ol{u},\wh{u},...)=0$. If we make the invertible gauge transformation
\beqn
\phi= G\psi 
\eeqn
where $G$ is an $N\times N$ matrix, then this Lax Pair becomes
\beqn\label{LP2}
\ol{G}\,\ol{\psi}=LG\psi, \qquad \wh{G}\wh{\psi}=MG\psi.
\eeqn
By Definition \ref{defFLP} this Lax pair is fake if we can choose a particular $G$ such that by using $Q(u,\ol{u},\wh{u},...)=0$ in \eqref{LP2}, the equations 
\beqn\label{LP3}
\ol{G}^{-1}LG=C, \hspace{1cm} \wh{G}^{-1}MG=D
\eeqn
holds for some constant matrices $C$ and $D$, where each element of $G$ is a closed-form expression involving $u$ and its various shifts, that is
\beqn\label{LPsol}
G_{ij}=G_{ij}(u,\ol{u},\wh{u},...).
\eeqn
The right hand sides of \eqn{LP3} are constant in this case because the elements of $L$ and $M$ are autonomous (except for their dependence on $u$). In general $C$ and $D$ may depend on the independent variables of the system, but must be independent of $u$ for the Lax pair to be fake. The Lax pair is therefore transformed to $\ol{\psi}=C\psi$, $\wh{\psi}=D\psi$. When $C=I$ the first equation in \eqref{LP3} has the formal solution
\beqn\label{LP4}
\ol{G}=L\ul{L}\,\ul{\ul L}...L_o
\eeqn
for some constant matrix $L_o$. Fake Lax pairs of type \eqref{LP1} have the special property that whenever we use $Q(u,\ol{u},\wh{u},...)=0$, the products $L\ul{L}$, $L\ul{L}\,\ul{\ul L}$ all simplify as algebraic expressions of $u$, $\ol{u}$, etc. such that \eqref{LP4} gives a closed-form expression for $G$. In practice, to find the simplest gauge transformation (usually when $C\neq I$) for fake Lax pairs of type \eqref{LP1}, one can use these products of $L$ to determine the location of all $u$'s within $G$, and choose the remaining constants in the simplest way such that $\det(G)\neq0$. An explicit example of this is shown in Example \eqref{example1}.
\newline\par
For Lax pairs of a different form to \eqref{LP1}, one can use similar techniques to find the required gauge transformation, as shown in the examples below.

\begin{remark} {\bf The Inverse Scattering Transform (IST) cannot be used on Fake Lax pairs}. The IST is a method of finding solutions to integrable nonlinear partial differential and difference equations, using their associated Lax pairs. In the 1+1 continuous case, given an initial condition $u(x,0)$, one first solves $\phi_x=L\phi$ (say), obtaining $\phi(x,0)$ and the scattering data as Neumann series in $u(x,0)$. The second Lax equation $\phi_t=M\phi$ then gives the simple time evolution of the scattering data, from which $\phi(x,t)$ (and thus $u(x,t)$) is obtained from a singular integral equation or Gelfand-Levitan-Marchenko (GLM) integral equation. For g-fake Lax pairs however one may gauge out ($\phi=G(u,u_x,...)\psi$) all dependent variables from the Lax pair and obtain $\psi(x,t)$ explicitly, for any initial condition. The time-dependent eigenfunction is then given by $\phi(x,t)=G(u(x,t),u_x(x,t),...)\psi(x,t)$, where the arguments of $G$ are still unknown, and are the very objects that we are trying to determine. Any attempt to use a singular integral or GLM equation for $\phi$ yields only identities such as $u(x,t)=u(x,t)$. In other words the eigenfunctions do not contain enough nontrivial information about the solution to be used for the IST. \end{remark}

\begin{remark}\label{rem:omega}
In the $2\times 2$ case, for autonomous Lax pairs of type \eqref{LP1}, one can perform a simple test to see if a diagonal gauge transformation
\beqn
\label{simpleG}
G=\left(\begin{array}{cc} g_1 & 0 \\ 0 & g_2 \end{array}\right)
\eeqn
exists that renders the Lax pair fake. Let $\left(\begin{array}{cc} A & B \\ C & D \end{array}\right)$ denote either $L$ or $M$, then from $\ol{G}^{-1}LG=const.$ and $\wh{G}^{-1}MG=const.$ one finds that if $A,B,C,D$ are all nonzero, such a gauge transformation exists only if
\beqn\label{dettest}
\frac{AD}{BC}=const.
\eeqn
If \eqref{dettest} holds for both $L$ and $M$, then the gauge transformation $G$ exists and the Lax pair is a fake. If \eqref{dettest} does not hold for both $L$ and $M$, then we must do more work to determine if a more complicated $G$ can be found. For the $2\times2$ continuous case, if the autonomous Lax pair $\phi_x=L\phi$, $\phi_t=M\phi$ satisfies
\beqn
BC=const., \quad \frac{\partial B}{B}=A-D+const.
\eeqn
for both $L$ and $M$, where $\partial=\partial_x$ for $L$ and $\partial_t$ for $M$, then a gauge transformation of the form \eqref{simpleG} may be found to remove all dependent variables, rendering the Lax pair a fake.

\end{remark}


\subsection{Method to directly identify excess freedom}\label{sec:freedom}
Given a Lax pair, the first step in applying the excess freedom test is to construct the generalised Lax pair of the same form. We can write the generalised Lax pair by maintaining the same dependence on the spectral parameter, and observing the coefficients of the linearly independent terms in the spectral parameter. Any of these coefficients that contain the dependent variable from the associated nonlinear system are replaced with arbitrary functions of the independent variables. The second step in the test is to find the form that these arbitrary functions must take,  as determined by the equations that arise from the compatibility condition.

\begin{definition}\label{defFLP2}
A Lax pair is called {\bf u-fake} if the set of equations yielded by the compatibility condition, operating on the generalised Lax pair of the same form, is underdetermined.
\end{definition}

If the system of equations resulting from the compatibility condition is underdetermined, we can use the excess freedom to associated any equation, integrable or not, with the Lax pair. Such a Lax pair must be fake since the alternative contradicts the property that the associated system is integrable, this notion is discussed in \cite{c01}. The excess freedom test is best illustrated with examples, see section \ref{sec:examples}.

\section{Examples}\label{sec:examples}
Many examples of fake Lax pairs arise in the literature, some were published as real but were later found to be fake, others were deliberately written as examples of fake Lax pairs. One expects that there are still others that are erroneously thought to be real and have gone unnoticed. In all cases of fake Lax pairs known to the authors, both of the tests presented in this article correctly identify them. Furthermore, it is impossible for either method, correctly applied, to wrongly identify a real Lax pair as fake.


\subsection{Example 1}\label{example1}
The following example is a fully discrete fake Lax pair that was originally found in \cite{h11}.
 
\subsubsection{Removal of dependent variable}
Here we give details of the gauge analysis (section \ref{sec:gauge}) that is used to remove the dependent variable. The Lax pair is
\beqn\label{ex1eq1}
\ol{\phi}=L\phi=\left(\begin{array}{cc} f & \ol{u}/\nu \\ 1/(\nu u) & 0 \end{array}\right)\phi, \qquad \wh{\phi}=M\phi=\left(\begin{array}{cc} \hat{u}/u & \nu \hat{u} \\ \nu/u & 1-\nu^2 \end{array}\right)\phi,
\eeqn
where $\nu$ is the spectral parameter and $u$ is the dependent variable of the nonlinear system. The compatibility condition of \eqref{ex1eq1} is
\beqn\label{ex1eq2}
f=\ol{u}/u,
\eeqn
where $f=f(u,\ol{u},...)$ can be {\it any} function of $u$ and its shifts. Since \eqref{ex1eq2} may be any equation whatsoever, it clearly should not have a real Lax pair. To show that \eqref{ex1eq1} is indeed fake, we use \eqref{ex1eq2} in \eqref{ex1eq1} to calculate
\[
L\ul{L}=\left(\begin{array}{cc} \ol{u}/\ul{u}(1+1/\nu^2) &\ol{u}/\nu \\ 1/(\nu \ul{u}) & 1/\nu^2 \end{array}\right), \qquad L\ul{L}\,\ul{\ul L}=\left(\begin{array}{cc} \ol{u}/\ul{\ul u}(1+2/\nu^2) & \ol{u}/\nu(1+1/\nu^2) \\ 1/(\nu \ul{\ul u})(1+1/\nu^2) & 1/\nu^2 \end{array}\right),
\]
which, as hinted by \eqref{LP4} because we first look for a local gauge that only depends on $u$ and not its shifts, suggests that $G$ will be of the form
\beqn\label{ex1eq3}
G=\left(\begin{array}{cc} ua & ub \\ c & d \end{array}\right),
\eeqn
for some constants $a$, $b$, $c$ and $d$. This is indeed the case, and the only restriction on these constants is that $\det(G)\neq0$, so the simplest choice is
\beqn\label{ex1eq3}
G=\left(\begin{array}{cc} u & 0 \\ 0 & 1 \end{array}\right).
\eeqn
Notice that $G$ is diagonal, which by Remark \ref{rem:omega} we could have obtained from the fact that for $M$ we have $\frac{AD}{BC}=const.$ The gauge \eqref{ex1eq3} transforms \eqref{ex1eq1} to
\beqn
\ol{\phi}=\left(\begin{array}{cc} 1 & 1/\nu \\ 1/\nu & 0 \end{array}\right)\phi \qquad \wh{\phi}=\left(\begin{array}{cc} 1 & \nu \\ \nu & 1-\nu^2 \end{array}\right)\phi.
\eeqn

\subsubsection{Excess freedom}
To write the generalised form of \eqn{ex1eq1} we maintain the dependence on the spectral parameter and replace any quantities that depend on $u$ with arbitrary functions of the independent variables. Thus we obtain
\beqn
L'= \left(\begin{array}{cc} a & b/\nu \\ c/\nu & 0 \end{array}\right), \qquad M'=\left(\begin{array}{cc} \al & \nu \be \\ \nu\ga & 1-\nu^2 \end{array}\right).
\eeqn
The compatibility condition, $\wh{L'}M'=\ol{M'}L'$ yields a set of equations for $a, b, c, \al, \be$ and $\ga$. If this set of equations is underdetermined, then the Lax pair is fake, which indeed it is in this case. The analysis was carried out in \cite{h11}, Appendix A, case 4, where $\de$ must be set to unity and the dependence on the spectral variable is different, but equivalent because it yields the same set of equations from the compatibility condition.

\subsection{Example 2} \label{sec:hone}
In this section we consider a Lax pair that was published as being real in \cite{h07}. The Lax pair is actually fake, but it was published as being associated with a nonlinear QRT mapping. The Lax pair is
\beqn\label{h07LP}
L=\left(\begin{array}{cc} u\bar{u} & -k_1 \\ \nu-u-\bar{u} & k_1(1/u+1/\bar{u})+k_2/(u\bar{u})\end{array}\right),\qquad M=\left(\begin{array}{cc} 0 & -k_1 \\ \nu-u-\bar{u} & k_1/\bar{u} \end{array}\right),
\eeqn
where $\nu$ is the spectral parameter, $k_1$ and $k_2$ are constants and $u$ is the dependent variable in the associated nonlinear system. The linear problem is
\beqn\label{eg2LP}
L\phi=\nu\phi,\qquad \bar{\phi}=M\phi
\eeqn
so the compatibility condition is $\ol{L}M=ML$, which associates the Lax pair with the following QRT mapping
\beqn
\bar{\bar{u}}=\frac{k_1 \bar{u}+k_2}{u\bar{u}^2}.\label{h07map}
\eeqn

\subsubsection{Removal of dependent variable}

An obvious property of \eqn{h07LP} is that
\beqn\label{eg2property}
L=M+D,
\eeqn
where $D$ is the diagonal matrix 
\beqn
D=\left(\begin{array}{cc} u\ol{u} & 0 \\ 0 & \ol{u}\,\ol{\ol u} \end{array}\right).
\eeqn
From \eqref{eg2LP} this implies
\beqn\label{eg2neweqn}
\bar{\phi}=(\nu I-D)\phi,
\eeqn
so we can set
\beqn\label{eg2gauge}
G=(\nu-u{\ol u})(\nu-{\ul u}u)(\nu-\ul{\ul u}\,{\ul u})\cdots(\nu-u_o)\left(\begin{array}{cc} 1 & 0 \\ 0 & \nu-u{\ol u} \end{array}\right)
\eeqn
which transforms \eqref{eg2neweqn} to $\ol{\psi}=\psi$. The Lax pair \eqref{h07LP} is therefore fake, as will be any Lax pair of the form \eqref{eg2LP} with the property $L=M+D$.

\subsubsection{Excess freedom}
The first thing to do is construct the generalised Lax pair with the same form in terms of the spectral parameter, $\nu$, but with arbitrary terms replacing the coefficients of the various powers of $\nu$ in each entry, including $\nu^0$, that depend on $u$. We arrive at
\beqn
L'=\left(\begin{array}{cc}a&-k_1\\\nu+c&d\end{array}\right),\qquad M'=\left(\begin{array}{cc}0&-k_1\\\nu+\ga&\de\end{array}\right),
\eeqn
where $\nu$ is the spectral parameter, $k_1$ is a constant and all other terms are arbitrary functions of the lattice variable $n$. 

The required form of the arbitrary terms is found by substituting the Lax pair into the compatibility condition, which is $\ol{L'}M'=M'L'$ in this case. Below we write out all the equations coming from the compatibility condition, separated into different powers of the spectral parameter, which is independent of other variables. This yields
\begin{align}
\ga&=c,& \bar{a}+\de&=d,& \bar{d}\ga&=a\ga+c\de,\label{andy1}\\
\bar{d}&=a+\de,&\bar{d}\de-k_1 \bar{c}&=d\de-k_1\ga.\label{andy2}
\end{align}

Now we solve this set of equations as follows. Allow (\ref{andy1}a) to define $\ga$, which causes (\ref{andy1}c) to coincide with (\ref{andy2}a).

Thus, the remaining compatibility conditions are:
\begin{align}
\bar{a}+\de&=d,\label{andya}\\
\bar{d}&=a+\de,\label{andyb}\\
b(\bar{c}-c)+\de(\bar{d}-d)&=0.\label{andyc}
\end{align}
The difference between equations \eqn{andya} and \eqn{andyb} implies that 
\beqn
d=k_3-a,\label{andyd}
\eeqn
where $k_3$ is a constant. Substituting this back into either \eqn{andya} or \eqn{andyb} gives us $\de=k_3-\bar{a}-a$. 

Now the only remaining compatibility condition to satisfy is \eqn{andyc}, substituting the above results brings this equation to
\beqn
k_1(\bar{c}-c)=\bar{a}^2-k_3\bar{a}-(a^2-k_3a), \label{cd1}
\eeqn
which is satisfied by
\beqn
c=\frac{a}{k_1}(a-k_3)+k_4, \label{cd}
\eeqn
where, $k_4$ is a constant..

We have now satisfied all of the elements of the compatibility condition. Collecting the results, we find that the Lax pair takes the following form, or one that is gauge equivalent:
\beqn
L'=\left(\begin{array}{cc}a&-k_1\\\nu+(a-k_3)a/k_1+k_4&k_3-a\end{array}\right),\qquad M'=\left(\begin{array}{cc}0&-k_1\\\nu+(a-k_3)a/k_1+k_4&k_3-\bar{a}-a\end{array}\right). \label{lp1}
\eeqn

If this was a real Lax pair, substituting it into the compatibility condition would yield an equation for $a$, which would be the associated nonlinear map. However, one can check that, in this case, the compatibility condition is identically satisfied and no conditions remain. This leaves $a$ completely free. This freedom proves that the Lax pair is fake and we can use it to write any arbitrary equation into the Lax pair.

As an example, we can retrieve an equivalent Lax pair to \eqn{h07LP} if, instead of allowing \eqn{andyd} to define $d$, we choose $d=k_1/u+k_1/\bar{u}+k_2/(u\bar{u})$, $a=u\bar{u}$ and set the $c$, $\ga$ and $\de$ according to the calculations given above with $k_4=k_2$. We can see that the resulting Lax pair is equivalent by noting that \eqn{h07map} can be summed to obtain the first order difference equation
\[
u\bar{u}+\frac{k_1}{u}+\frac{k_1}{\bar{u}}+\frac{k_2}{u\bar{u}}=k_3,
\]
where $k_3$ is a constant. Using \eqn{h07map} to replace $k_1/u+k_2/(u\bar{u})$ in this expression shows that $u$ must also satisfy
\[
u\bar{u}+\bar{u}\bar{\bar{u}}+\frac{k_1}{u\bar{u}}=k_3,
\]
It follows that the two Lax pairs, both fake, are equivalent.


\subsection{Example 3} \label{sec:gramani}
In this section we analyse a fake Lax pair that was published as a real one in \cite{gr04} (equations (3.23) in that paper). The Lax pair is
\beqn\label{eg3LP}
\phi(q\nu)=L\phi(\nu), \qquad \bar{\phi}(\nu)=M\phi(\nu)
\eeqn
where
\beqn\label{eg3LPmat}
L=\left(\begin{array}{cccc} 0 & 0 & k/u & 0 \\ 0 & 0 & \ul{u} & q\ul{u} \\ \nu u & 0 & 1 & q \\ 0 & \nu \ul{k}/\ul{u} & 0 & 0 \end{array}\right) \qquad M=\left(\begin{array}{cccc} 0 & u/(k(u+1)) & 0 & 0 \\ 0 & 0 & 1 & 0 \\ 0 & 0 & 1/u & q/u \\ \nu & 0 & 0 & 0 \end{array}\right)
\eeqn
and $k$ is a function of $n$ such that $\bar{\bar{k}}=qk$. The compatibility $\ol{L}(\nu)M(\nu)=M(q\nu)L(\nu)$ yields $q$-P$_I$
\beqn\label{qP1}
\ul{u}\bar{u}=k\bar{k}\left(\frac{1+u}{u^2}\right).
\eeqn

\subsubsection{Removal of dependent variable}

We look for a diagonal gauge transformation. Define the function $f_n$ by 
\beqn
f_n=\prod_{i=i_o}^{n-1}1/u_i,
\eeqn
and let the gauge transformation $G$ be
\beqn\label{eg3gauge}
G=\left(\begin{array}{cccc} \bar{f}/q & 0 & 0 & 0 \\ 0 & \ul{f} & 0 & 0 \\ 0 & 0 & f & 0 \\ 0 & 0 & 0 & f/q \end{array}\right).
\eeqn
Using \eqref{qP1} we then find that
\beqn
G^{-1}LG=\left(\begin{array}{cccc} 0 & 0 & qk & 0 \\ 0 & 0 & 1 & 1 \\ \nu/q & 0 & 1 & 1 \\ 0 & \nu \bar{k} & 0 & 0 \end{array}\right), \qquad \ol{G}^{-1}MG=\left(\begin{array}{cccc} 0 & q\bar{k} & 0 & 0 \\ 0 & 0 & 1 & 0 \\ 0 & 0 & 1 & 1 \\ \nu & 0 & 0 & 0 \end{array}\right),
\eeqn
and so the Lax pair \eqref{eg3LPmat} is transformed to
\beqn
\psi(q\nu)=\left(\begin{array}{cccc} 0 & 0 & qk & 0 \\ 0 & 0 & 1 & 1 \\ \nu/q & 0 & 1 & 1 \\ 0 & \nu \bar{k} & 0 & 0 \end{array}\right)\psi(\nu), \qquad \bar{\psi}(\nu)=\left(\begin{array}{cccc} 0 & q\bar{k} & 0 & 0 \\ 0 & 0 & 1 & 0 \\ 0 & 0 & 1 & 1 \\ \nu & 0 & 0 & 0 \end{array}\right)\psi(\nu),
\eeqn
showing that it is indeed fake.

\subsubsection{Excess freedom}
To show that the excess freedom exists in the compatibility condition of this Lax pair, we first write a generalised form of the Lax pair with the same dependence on the spectral parameter. Noting \eqn{eg3LPmat} we immediately write the generalised form
\beqn
L'=\left(\begin{array}{cccc} 0&0&a&0\\0&0&b&c\\\nu u&0&1&q\\0&\nu d&0&0\end{array}\right),\qquad 
M'=\left(\begin{array}{cccc} 0&\al&0&0\\0&0&1&0\\0&0&\be&\ga\\ \nu&0&0&0\end{array}\right),
\eeqn
where $\nu$ is the spectral parameter and all other variables inside the Lax matrices are arbitrary functions of the discrete lattice variable $n$. Notice that we have left the dependent variable $u$ in the (3,1) entry of $L'$, but at this stage we do not know what equation $u$ must satisfy, so essentially it is an arbitrary function of the independent variable, $n$, and this is the generalised Lax pair.

Substituting these matrices into the compatibility condition $L'(q\nu)M'=\ol{M'}L'$ yields the following set of equations
\begin{align}
\bar{a}\be&=b\al,&\bar{a}\ga&=c\al,&\bar{b}\be&=1,\label{gr1}\\
\bar{b}\ga&=q,&u\be&=q,&\bar{u}\al&=d\ga,\label{gr2}\\
\ga&=q\be,&\bar{d}&=qa.\label{gr3}
\end{align}
Thus, the compatibility condition delivers a set of eight equations for the eight unknown functions, and all may appear to be well, but some equations coincide leaving excess freedom in this system. Naming the equations in \eqn{gr1} as (\ref{gr1}a), (\ref{gr1}b) and (\ref{gr1}c), respectively (and naming \eqn{gr2} and \eqn{gr3} the same way), we do the following. Allow (\ref{gr2}b), (\ref{gr1}c), (\ref{gr3}a) and (\ref{gr3}b) to define $\be$, $b$, $\ga$ and $d$, respectively. Notice that these definitions cause (\ref{gr2}a) to coincide with (\ref{gr1}c). Thus, the remaining equations become
\begin{align}
q^2\bar{a}/u&=\al\ul{u},&q^2\bar{a}/u&=c\al,&\bar{u}\al&=q^3\ul{a}/u.\label{gr4}
\end{align}
Comparing (\ref{gr4}a) and (\ref{gr4}b) shows that $c=\ul{u}$. Now use (\ref{gr4}c) to define $a$, which causes the final remaining equation, (\ref{gr4}a) to become
\beqn
\bar{\bar{\bar{u}}}\bar{\bar{u}}\bar{u}\bar{\bar{\al}}=q\bar{u}u\ul{u}\al.\label{grlinear}
\eeqn

\begin{remark}
At this point in the analysis, one should make a change of variables such as $\al=x/(\bar{u}u\ul{u})$ so that \eqn{grlinear} becomes $\bar{\bar{x}}=qx$. The Lax pair is real when associated with this trivial equation for $x$. However, we can construct fake Lax pairs by continuing without this change of variables.
\end{remark}

Naturally, \eqn{grlinear} is solved by
\[\al=\frac{k}{\bar{u}u\ul{u}},\]
where $\bar{\bar{k}}=qk_n$ and there is one excess degree of freedom. This freedom can be used to write a fake Lax pair for any equation, writing $\al=u/k(1+u)$ returns the fake Lax pair for $q$-P$_I$ from \cite{gr04}.

For the reader's convenience, the full fake Lax pair is written below:
\beqn
L'=\left(\begin{array}{cccc} 0&0&\bar{\bar{u}}\bar{u}\bar{\ala}/q^3&0\\0&0&\ul{u}/q&\ul{u}\\\nu u&0&1&q\\0&\nu \bar{u}u\al/q^2&0&0\end{array}\right),\qquad 
M'=\left(\begin{array}{cccc} 0&\al&0&0\\0&0&1&0\\0&0&q/u&q^2/u\\ \nu&0&0&0\end{array}\right),
\eeqn
where $\al$ is arbitrary.

\subsection{Example 4}

As a final example we consider the fake Lax pair
\beqn\label{eg4LP}
\phi_x=\left(\begin{array}{cc} 0 & 1 \\ \lambda f^2 & \mu f +f_x/f \end{array}\right)\phi, \qquad \phi_t=\left(\begin{array}{cc} \nu & (g+\rho)/f \\ \lambda f(g+\rho) & \nu+\mu(g+\rho)+g_x/f \end{array}\right)\phi
\eeqn
which was given in \cite{cn91} and identified as fake in that paper. Note that the Lax pair was given in scalar form in \cite{cn91}, equation (6), but transforming it into matrix form is straightforward, as mentioned in \cite{s01}. The compatibility condition of \eqref{eg4LP} is
\beqn\label{eg4compat}
f_t=g_x
\eeqn
for any functions $f$ and $g$, and $\lambda, \mu, \nu, \rho$ are all independent parameters. This fake Lax pair can be used to represent a large class of nonlinear systems by arbitrarily writing $f$ and $g$ in terms of $u$, which becomes the dependent variable in the nonlinear system.

\subsubsection{Removal of dependent variable}
Consider first the diagonal gauge transformation
\beqn
G=\left(\begin{array}{cc} 1 & 0 \\ 0 & f \end{array}\right),
\eeqn
which, using \eqref{eg4compat}, transforms \eqref{eg4LP} to
\beqn\label{eg4LPa}
\psi_x=f\left(\begin{array}{cc} 0 & 1 \\ \lambda & \mu \end{array}\right)\psi, \qquad \psi_t=\nu\psi+(g+\rho)\left(\begin{array}{cc} 0 & 1 \\ \lambda & \mu \end{array}\right)\psi.
\eeqn
If we let $A=\left(\begin{array}{cc} 0 & 1 \\ \lambda & \mu \end{array}\right)$ then we can make the second gauge transformation
\beqn
\psi=\exp\left(A\int(g+\rho)dt+\nu t\right)\psi^{o},
\eeqn
which using \eqref{eg4compat} transforms \eqref{eg4LPa} to $\psi^o_x=\psi^o_t=0$. Thus the combined gauge transformation $\phi=G\exp\left(A\int(g+\rho)dt+\nu t\right)\psi^{o}$ removes all dependent variables from the Lax pair \eqref{eg4LP}.

\subsubsection{Excess freedom}
In fact, this Lax pair fails the excess freedom test by construction, since the compatibility condition is associated with an underdetermined system in \eqn{eg4compat}. We go ahead with the analysis anyhow, to see how to analyse any case where, for example, $g$ is given in terms of $f$ (or $f$ and $g$ given in terms of some $u$), so that the Lax pair's fakeness is not obvious. A generalised form of the Lax pair in \eqn{eg4LP} is given by
\[
L'=\left(\begin{array}{cc}0&1 \\ \la c & \mu d + b\end{array}\right),\qquad M'=\left(\begin{array}{cc}\nu&\be \\ \la \ga & \nu  + \mu \de + \al\end{array}\right).
\]
Here the extraneous parameter $\rho$ has been excluded because it is inconsequential and the introduced variables $b$, $c$, $d$, $\al$, $\be$, $\ga$ and $\de$ all depend on both the continuous independent variables, $x$ and $t$. The remaining quantities, $\la$, $\mu$ and $\nu$ can all be considered spectral parameters.

The compatibility condition is $L'_t+L'M'=M'_x+M'L'$ from which we obtain the following set of equations
\begin{align}
\ga&=c\be,&\de&=d\be,&d\ga&=c\de,\label{c1}\\
d_t&=\de_x,&b_t&=\al_x,&\al&=\be_x+b\be,\label{c2}\\
c_t+b\ga&=\ga_x+c\al.\label{c3}
\end{align}
We solve this set of equations as follows: let (\ref{c1}a) and (\ref{c1}b) define $\ga$ and $\be$ respectively, this causes (\ref{c1}c) to become an identity. Let (\ref{c2}c) define $\al$ and substitute all these definitions into (\ref{c2}b) and (\ref{c3}), which become
\begin{align}
b_t&=\left[\left(\frac{\de}{d}\right)_x+\frac{b\de}{d}\right]_x,\label{eg4:b}\\
c_t&=c_x\frac{\de}{d}+2c\left(\frac{\de}{d}\right)_x,\label{eg4:c}
\end{align}
respectively. Equation \eqn{eg4:b} is solved by introducing $u=u(x,t)$ such that
\[b=u_x,\qquad \left(\frac{\de}{d}\right)_x+\frac{b\de}{d}=u_t.\]
Using (\ref{c2}a) shows that the latter of these is satisfied when $u=\log{d}$. Once again using (\ref{c2}a), we find that \eqn{eg4:c} is satisfied when $\log{c}=2\log{d}$. 

Now the system of equations obtained from the compatibility condition is satisfied. There are various ways to write the resulting Lax pair, allowing $d=f$, $\de=g$ and expressing the other variables in terms of these retrieves \eqn{eg4LP} (with $\rho=0$). The system of equations is undetermined, which confirms that the Lax pair is fake. We could take any of the compatibility conditions to be the associated nonlinear system, in \cite{cn91} they choose (\ref{c2}a) to play that role.

\section{Conclusion}\label{sec:conclusion}

In this letter, we have provided two very simple methods to identify fake Lax pairs. The methods have been used to confirm that several Lax pairs taken from the literature are fake, some of which were intentionally given as fake, while others were thought to be real. 

Both methods are widely applicable, with the only restriction being that the Lax par should contain a non-removable spectral parameter. We leave as open questions any discussion about the usability and reality of Lax pairs that do not contain a spectral parameter and do not allow for one to be inserted. However, it is the authors' opinion that such Lax pairs are also fake, since they do not allow for any spectral analysis.

The extent to which a fake Lax pair can be used to gain information about its associated system is also open to debate. For example in \cite{k80} a Lax pair for Burger's equation is given. The Lax pair is fake according to the tests given here and it cannot be used for inverse scattering. However, the Lax pair is related to a Cole-Hopf transformation which linearises Burger's equation and is thus useful. We do not discuss it further here, but some sections of \cite{c01} are devoted to this issue.


\end{document}